\documentclass[twocolumn]{aastex63}
\usepackage{lineno}
\usepackage[caption=false]{subfig}
\usepackage{multirow}
\usepackage{graphicx}
\usepackage{booktabs}
\usepackage{threeparttable}
\usepackage{amsmath,amssymb}
\usepackage[T1]{fontenc}
\usepackage{CJKutf8}
\usepackage{placeins}

\graphicspath{{./}{figures/}}

\begin{document}
\begin{CJK}{UTF8}{gbsn}

\title{Searching for $\gamma$-ray emission from a bona fide Compact Symmetric Object sample: a $\gamma$-ray signal near GB6 J0906+4124}

\correspondingauthor{Ying-Ying Gan}
\email{Yingying\_Gan@163.com}

\author[0000-0002-4789-7703]{Ying-Ying Gan\dag}
\affiliation{Information Support Force Engineering University, Wuhan 430000, People's Republic of China; Yingying\_Gan@163.com}

\author[0000-0002-5582-8265]{Kai Wang}
\affil{School of Astronomy and Space Science, Nanjing University, Nanjing 210023, Jiangsu, China}

\author[0000-0003-2547-1469]{Ji-Shun Lian}
\affiliation{School of Physics, Beijing Institute of Technology, Beijing 100081, People's Republic of China}

\author{Run-Meng Wang}
\affiliation{School of Physics, Beijing Institute of Technology, Beijing 100081, People's Republic of China}

\begin{abstract}

As a particular subclass of active galactic nuclei (AGNs), compact symmetric objects (CSOs) have attracted significant attention due to potential role as young AGNs. 
Several $\gamma$-ray emitting CSOs have been detected with Fermi Large Area Telescope (Fermi-LAT), which motivates further searches for more $\gamma$-ray emitting CSOs. 
We perform a systematic search for $\gamma$-ray emission from a bona fide CSO sample using 16 yr Fermi-LAT observation data. No significant $\gamma$-ray signal is found to be firmly associated with any CSO. Only one $\gamma$-ray signal is detected near CSO GB6 J0906+4124 with TS = 28.7 ($\sim4.9\sigma$) in 0.1--300 GeV band. 
Within the 68\% containment radius of Fermi-LAT average PSF at 3 GeV, GB6 J0906+4124 remains the plausible counterpart of the $\gamma$-ray signal, primarily due to its classification and broad spectral coverage.
This work suggests that GeV emission from CSOs may either be uncommon or simply below the current Fermi-LAT detection sensitivity. Further multi-wavelength follow-up is needed to determine the origin of the $\gamma$-ray signal near GB6 J0906+4124.

\end{abstract}


\section{Introduction}

Compact symmetric objects (CSOs) are a special subclass of active galactic nuclei (AGNs), initially distinguished by compact radio structures with a projection size of $\leqslant$1 kiloparsec (kpc) and two-sided symmetric jets (\citealp{1980ApJ...236...89P, 1994ApJ...432L..87W, 1996ApJ...460..634R}).
Based on radio structure, CSOs can be divided into two classes: edge-brightened (lobe-dominated) CSOs and edge-dimmed (core-dominated) CSOs (\citealp{2024ApJ...961..242R}).
CSO jet axes lie close to the plane of the sky (\citealp{1996ApJ...460..612R,2000ApJ...534...90P,2003ApJ...597..157T}), so that the observed radiation does not exhibit a strong beaming effect (\citealp{1980ApJ...236...89P, 1994ApJ...432L..87W, 1996ApJ...460..634R}). For instance, CSO OQ +208 has been measured to have an inclination angle of $\sim80.8^\circ$ between its radio jet and the line of sight (\citealp{2013A&A...550A.113W}). CSOs are considered young, jet-obstructed, and/or short-lived AGNs, which typically results in a miniature radio structure (\citealp{1980ApJ...236...89P, 1984AJ.....89....5V, 1995A&A...302..317F, 1997ASPC..121..672B, 1996ApJ...460..634R, 2017ApJ...836..174C, 2024ApJ...961..242R}). Recently, to ensure the purity of the CSO population, \cite{2024ApJ...961..240K} added two new classification criteria for CSOs, i.e., long-term low variability with a fractional variability of < 20\% yr$^{-1}$ and low apparent speeds of $\nu_{\rm app} < 2.5c$ along the jets, and then compiled a bona fide CSO catalog which contains 79 sources. 

The origin of the $\gamma$-ray emission in CSOs is still debated.
Theoretically, the mini-lobes of CSOs can produce GeV emission via leptonic inverse-Compton (IC) processes (\citealp{2008ApJ...680..911S, 2010ApJ...715.1071O}) or through hadronic processes (\citealp{2011MNRAS.412L..20K}).  The high-energy emission in CSOs could also be due to IC processes in mini-jets (\citealp{2014ApJ...780..165M}).
Alternatively, thermal bremsstrahlung radiation from the cocoon has been proposed as a possible mechanism for MeV-GeV emission in these sources (\citealp{2007MNRAS.376.1630K, 2009MNRAS.395L..43K}). In addition, the shell of these sources may also contribute to the high-energy emission via IC scattering (\citealp{2011ApJ...730..120I}).
In the 79 CSOs, TXS 0128+554 (\citealp{2020ApJ...899..141L, 2024RAA....24b5018G}), NGC 3894 (\citealp{2020A&A...635A.185P, 2024RAA....24b5018G}), and PKS 1718--649 (\citealp{2016ApJ...821L..31M, 2024RAA....24b5018G}) have been detected in $\gamma$-ray band with Fermi Large Area Telescope (Fermi-LAT); their low luminosity and non-variable $\gamma$-ray radiation is believed to originate from the IC process of electrons in jet extended regions (i.e., lobes) based on multi-band spectral energy distribution (SED) modeling (\citealp{2024RAA....24b5018G, 2024A&A...684A..65B}). It should be noted that NGC 4278 is also contained in the bona fide CSO catalog, which has been reported to be associated with a very high energy (VHE) $\gamma$-ray source 1LHAASO J1219+2915 and exhibits variability on radio, X-ray, and TeV bands (\citealp{2005ApJ...622..178G, 2024ApJ...974..134L, 2024ApJ...971L..45C, 2024ApJS..271...25C}). Its $\gamma$-ray emission detected with Fermi-LAT has been reported in \cite{2024ApJ...977L..16B}. Multi-band SED modeling reveals that the $\gamma$-ray emission of NGC 4278 likely originates from non-thermal radiation produced via the IC scattering of electrons in the core jet (\citealp{2024ApJ...974..134L, 2024ApJ...967..137W, 2024ApJ...974...56D}). 
Another CSO OQ +323 (DA 362) is also included in the bona fide CSO catalog. And, it was reported to be spatially coincident with a $\gamma$-ray source 4FGL J1416.0+3443 and was defined as a blazar candidate of uncertain type in 4FGL-DR4 (\citealp{2023arXiv230712546B}). A transient $\gamma$-ray activity from this source was reported by \cite{2021ApJS..256...13B}. Moreover, \cite{2025ApJ...979...97S} has recently reported the $\sim 15.75$ yr Fermi-LAT observation of OQ +323. 
Besides, a $\gamma$-ray emitting source 4C +39.23B (\citealp{2022ApJ...927..221G}) is also included in the bona fide CSO catalog, but its total projected linear size has been reported to be greater than 1 kpc and was previously classified as a compact steep-spectrum source (\citealp{2004A&A...426..463O}). 

$\gamma$-ray emitting CSOs are still rare. 
In addition to the several $\gamma$-ray emitting CSOs mentioned above, there are other two $\gamma$-ray emitting CSOs that were not listed in the bona fide CSO catalog due to the flux variability similar to blazar, i.e., CTD 135 (\citealp{2022Symm...14..321F}) and PKS 1413+135 (\citealp{2024ApJ...961..240K}).
To expand $\gamma$-ray emitting CSO sample and further study the $\gamma$-ray radiation properties of CSOs, we use 16 yr Fermi-LAT observation data to search for GeV signals from CSOs. The structure of this paper is as follows: Section 2 explains the selection of CSO sample. A description of the Fermi-LAT data analysis and results is presented in Section 3. Section 4 is the discussion. Finally, a summary is presented in Section 5. Throughout, $H_0=71$ km s$^{-1}$ Mpc$^{-1}$, $\Omega_{\rm m}=0.27$, and $\Omega_{\Lambda}=0.73$ are adopted in this paper.

\label{sec:intro}

\section{Sample Selection}

\cite{2024ApJ...961..240K} has compiled a bona fide CSO catalog including 79 sources, in which the $\gamma$-ray signals from 6 sources have been confirmed, i.e., TXS 0128+554 (\citealp{2020ApJ...899..141L, 2024RAA....24b5018G}), NGC 3894 (\citealp{2020A&A...635A.185P, 2024RAA....24b5018G}), PKS 1718--649 (\citealp{2016ApJ...821L..31M, 2024RAA....24b5018G}), OQ +323 (\citealp{2025ApJ...979...97S}), NGC 4278 (\citealp{2024ApJ...977L..16B}), and 4C +39.23B (\citealp{2022ApJ...927..221G}). In this work, we use the remaining 73 sources as a sample to search new $\gamma$-ray emitting CSO. Table 1 lists the names and redshifts of the 73 CSOs, taken from \cite{2024ApJ...961..240K} and NASA/IPAC Extragalactic Database (NED; \citealp{NED})\footnote{https://ned.ipac.caltech.edu/}. As seen in the Table 1, the redshifts of 57 sources are known, the lowest and highest redshifts among these CSOs are 0.01 for GB6 J1204+5202 and 1.975 for B21225+36, respectively.

\section{Fermi-LAT Data Analysis and Results}

\subsection{Data Selection}

The Fermi-LAT on board the Fermi satellite is a converter telescope capable of detecting $\gamma$-rays from 20 MeV to > 300 GeV (\citealp{2009ApJ...697.1071A}), making it a powerful instrument for the long-term monitoring of AGNs.
We extract Fermi-LAT Pass 8 observation data within a $15^{\circ}$ region of interest (ROI) from the Fermi Science Support Center, centered at the positions of the 73 candidate sources provided by \cite{2024ApJ...961..240K}.
The data set spans approximately 16 yr, from 2008 August 4 to 2024 August 9 (MJD 54682--60531). Photon events within the 100 MeV to 300 GeV energy range are considered in the data analysis. To mitigate the effects of the background $\gamma$-ray contamination from the Earth limb, we retain only photon events with zenith angles $\leqslant 90^{\circ}$. The standard data quality selection criteria ``(DATA\_QUAL$>$0)$\&\&$(LAT\_CONFIG==1)'' ensures the selection of high-quality photon events.
We bin the data using a pixel size of $0.1^{\circ}$ and 12 logarithmic energy bins.
The instrument response function P8R3\_SOURCE\_V3 is employed in the data analysis. The public software \textit{fermitools}\footnote{https://fermi.gsfc.nasa.gov/ssc/data/analysis/software/} (ver. 2.2.0; \citealp{2019ascl.soft05011F}) and python package \textit{Fermipy}\footnote{http://fermipy.readthedocs.io/en/latest/} (ver. 1.1.1; \citealp{2017ICRC...35..824W}) with the binned likelihood analysis method are adopted to analyze the data. 

\subsection{TS Test}

The background model for each celestial ROI is composed of the diffuse Galactic interstellar emission (gll\_iem\_v07.fits), the isotropic emission (iso\_P8R3\_SOURCE\_V3\_v1.txt), and all point sources reported in the Fermi-LAT 14-yr Source Catalog (4FGL-DR4, \citealp{2023arXiv230712546B}). 
The maximum test statistic (TS) is adopted to quantify the significance of the $\gamma$-ray detection. TS is defined as $2{\rm log}(\frac{\mathcal{L}_{1}}{\mathcal{L}_{0}})$ (\citealp{1996ApJ...461..396M}), 
which compares the likelihood between a model containing the point source and a model without it.
Significant $\gamma$-ray excess with TS $\geqslant$ 25 indicates a source detection. 
We optimize the model fitting in the ROI.
We then generate TS residual maps centered at the positions of the 73 candidate sources to determine whether any faint sources not listed in the 4FGL-DR4 appear nearby.
For individual 4FGL-DR4 source, we assume that its spectral model and corresponding parameters are the form and values reported in the 4FGL-DR4, respectively. The spectral parameters of 4FGL-DR4 sources within $6^{\circ}$, as well as the normalization of the diffuse Galactic interstellar emission and the isotropic emission, are left free. 
The candidate sources are modeled by point sources with power-law spectra, i.e.,
\begin{equation}
\frac{dN}{dE} = N_0(\frac{E}{E_0})^{-\Gamma_{\gamma}},
\end{equation}
at the corresponding positions provided by \cite{2024ApJ...961..240K}, where $N_0$ is the normalization parameter, $\Gamma_{\gamma}$ is the photon spectral index, and $E_0$ is the scale parameter.

Based on the above standard data analysis method, we detect only one $\gamma$-ray signal with TS $\geqslant$ 25, located near GB6 J0906+4124. Three sources have TS values between 10 and 25, while the remaining sources have TS values below 10. Given this, we perform a more detailed analysis for GB6 J0906+4124. And the remaining sources continue to use standard data analysis method.
For sources with TS $\geqslant$ 10, the $N_0$ and $\Gamma_{\gamma}$ are left free to vary, and we derive the 0.1--300 GeV photon flux of the sources. For sources with TS < 10, we only derive the 0.1--300 GeV photon flux upper limits at 95\% confidence level by fixing $\Gamma_{\gamma}$ to 2.0.

\subsection{Data Analysis Results for CSO GB6 J0906+4124}

\begin{figure*}
    \centering
    \includegraphics[angle=0, scale=0.42]{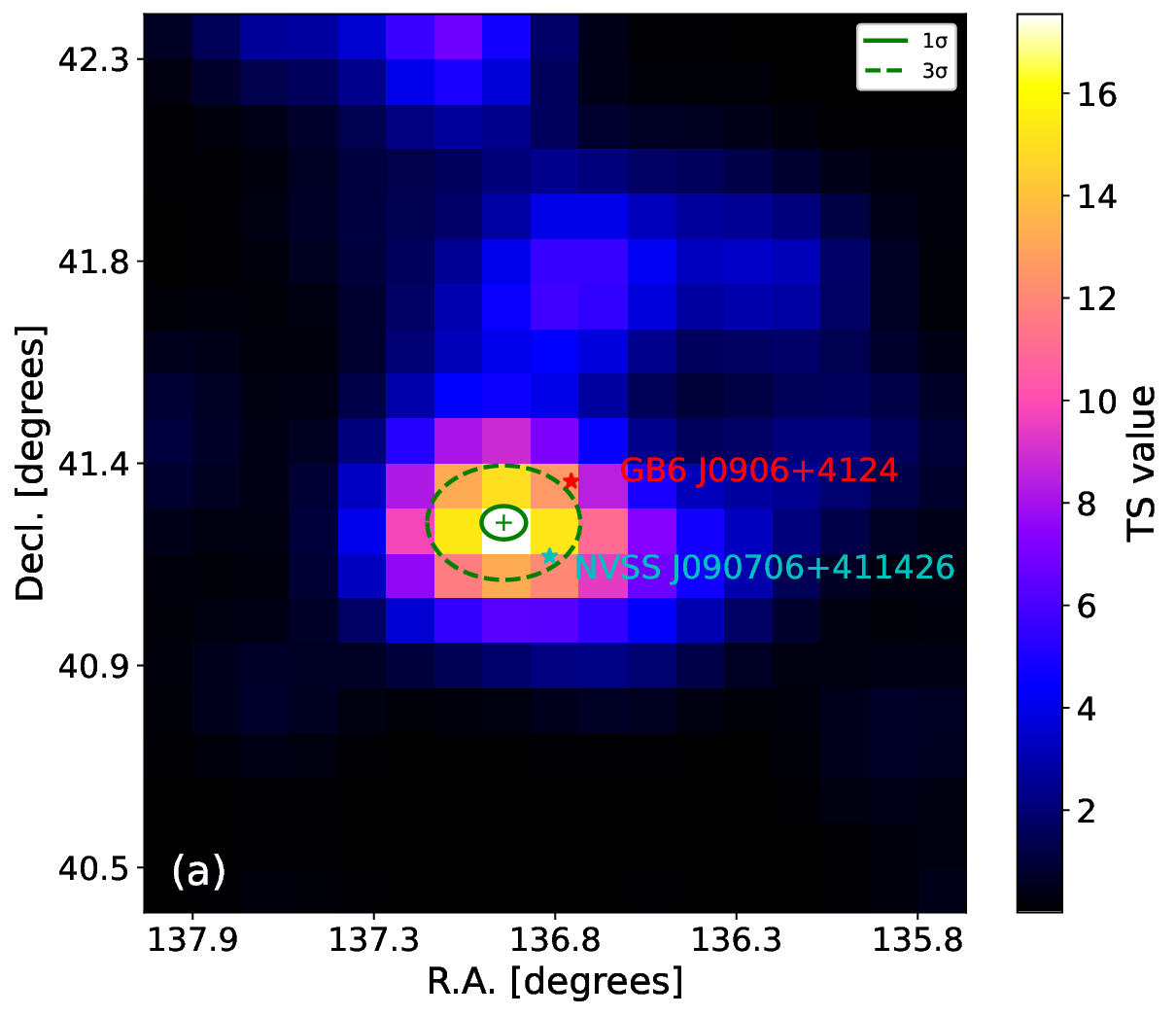}
    \includegraphics[angle=0, scale=0.42]{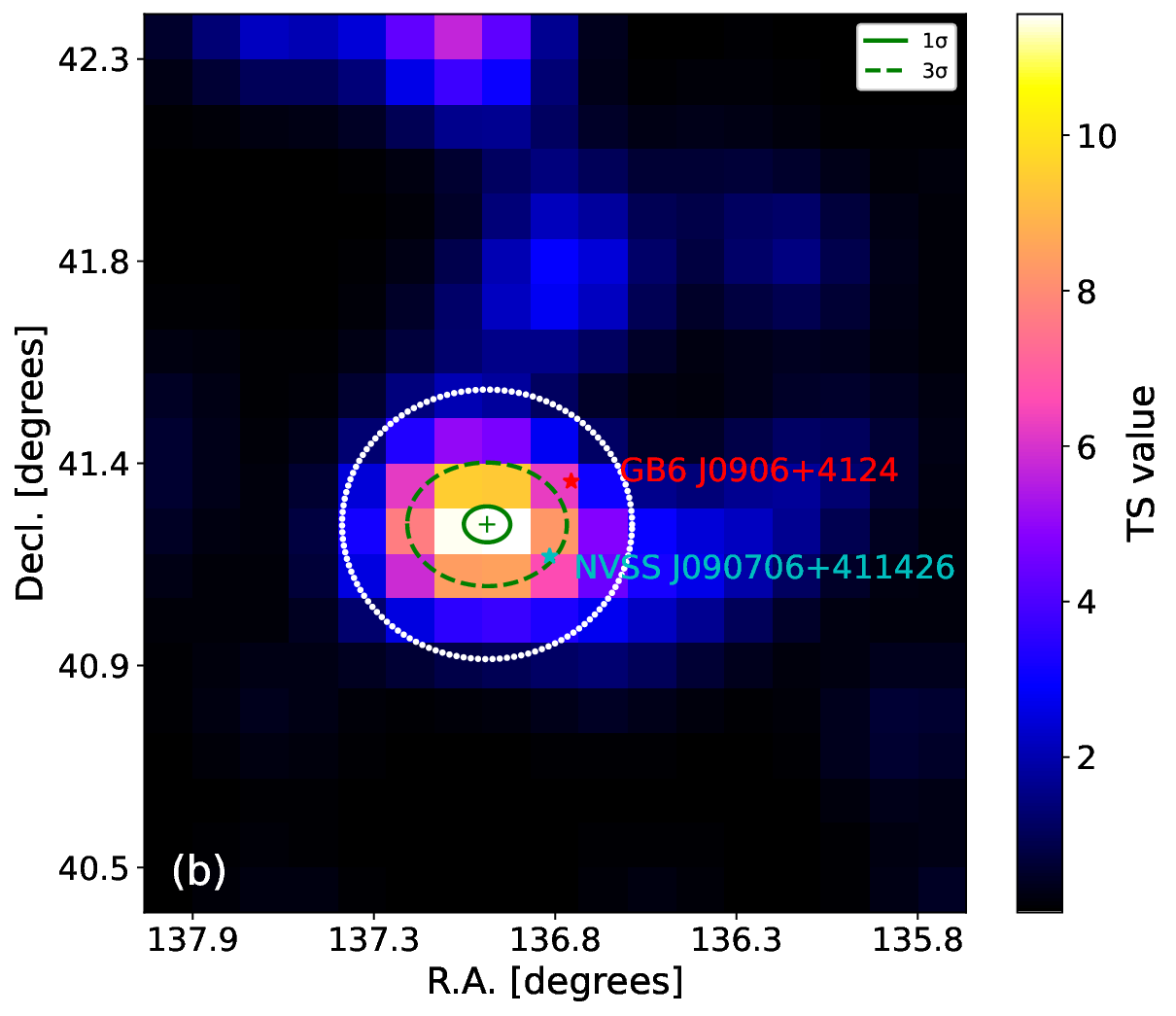}
    \caption{Panel (a): $2.0\degr \times 2.0\degr$ TS map in the 0.1--300 GeV band for the new $\gamma$-ray signal. The green symbols indicate the best-fit position (green cross), $1\sigma$ (green solid ellipse) and $3\sigma$ (green dashed ellipse) uncertainty ellipse based on 16 yr Fermi-LAT observation data. The red and cyan star symbols represent the positions of CSO GB6 J0906+4124 and radio source NVSS J090706+411426. Panel (b): same as the Panel (a), but in the 3--300 GeV band. And, the white dotted circle indicates the 68\% containment radius of the Fermi-LAT average PSF at 3 GeV ($\sim 0.3\degr$; \citealp{2009ApJ...697.1071A}), which is centered at the best-fit position of the new $\gamma$-ray signal. The TS maps are generated with a pixel size of $0.1\degr$ and are smoothed.}
    \label{TSmap}
\end{figure*}

\begin{figure*}
    \centering
    \includegraphics[angle=0, scale=0.5]{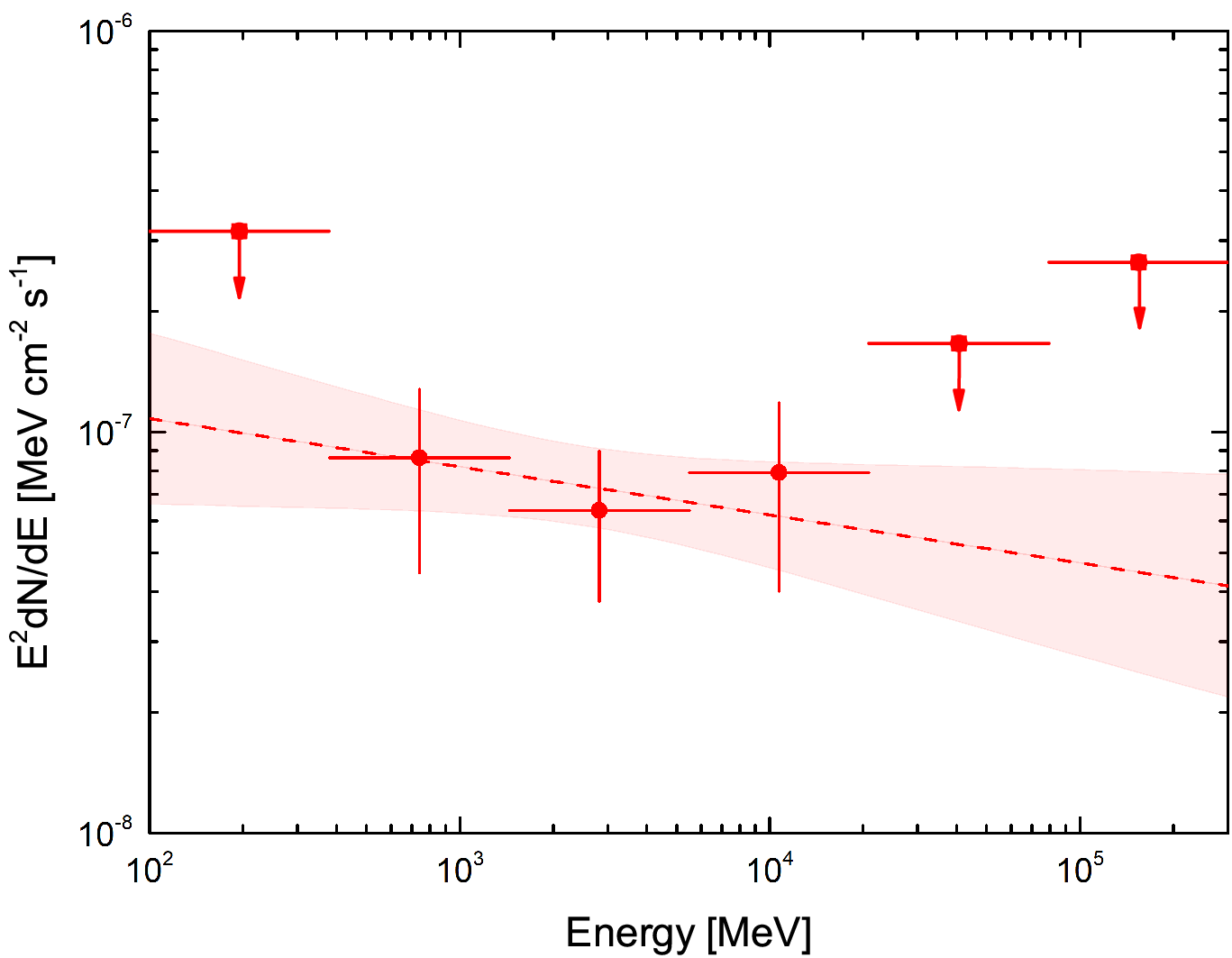}
    \caption{The 0.1--300 GeV average spectrum of the new $\gamma$-ray signal. The red dashed line represents the fitting result with power-law model, and the red shadow represents the $1\sigma$ error band. The threshold of TS = 4 for an energy bin is applied, i.e., an upper limit is presented if TS < 4.}
    \label{spectrum}
\end{figure*}

Based on the preceding data analysis, we discover a $\gamma$-ray signal close to GB6 J0906+4124. We then perform a more detailed analysis, following the procedure adopted in \cite{2022ApJS..260...53A}. For energies between 100 MeV and 300 MeV we exclude events with zenith angle larger than 90 , as well as photons from PSF0 and PSF1 event types, while between 300 MeV and 1 GeV we exclude events with zenith angle larger than 100, as well as photons from the PSF0 event type. Above 1 GeV we use all events with zenith angles less than 105. 
We perform an independent search for new $\gamma$-ray sources using \textit{find\_sources}( ) across the entire ROI. We still detect a $\gamma$-ray signal near GB6 J0906+4124, and two additional new $\gamma$-ray signals are also found. These two additional $\gamma$-ray signals are included as background sources in our analysis.
The 0.1--300 GeV TS map of the $\gamma$-ray signal near GB6 J0906+4124 is generated, as shown in Figure \ref{TSmap}(a). Using \textit{Fermipy} package with the 16 yr Fermi-LAT 0.1--300 GeV observation data, we estimate the best-fit position of the $\gamma$-ray signal. The best-fit position is [R.A.= 136.905$\degr$, Decl.= 41.315$\degr$], the position angle of uncertainty ellipse is 179.841$\degr$, the corresponding $1\sigma$ uncertainty along major axis and minor axis of uncertainty ellipse are 0.046$\degr$ and 0.037$\degr$, respectively. We plot the best-fit position and the corresponding $1\sigma$ and $3\sigma$ error ellipses in Figure \ref{TSmap}(a). The radio position of GB6 J0906+4124 is 0.167$\degr$ away from the best-fit position of the $\gamma$-ray signal and is not within $3\sigma$ error ellipse, as seen in Figure \ref{TSmap}(a). 
The TS value of the $\gamma$-ray signal is 28.7 ($\sim4.9\sigma$)\footnote{TS = 28.7 with 2 degrees of freedom is obtained before the best-fit localization, corresponds to an estimated statistical significance of $\sim4.9\sigma$ (see \citealp{1996ApJ...461..396M}). Note that this is a local significance, following the formalism reported in Section 7 of  \cite{2012A&A...540A..79B} (in the special case of $k=1$ and $M=1$), the global significance, corrected for 73 independent trials, is $\sim4.0\sigma$.}. 
We obtain $\Gamma_{\gamma} = 2.1\pm0.2$ and photon flux $F_{\rm 0.1-300 GeV} = (9.7\pm5.6) \times 10^{-10}$ ph cm$^{-2}$ s$^{-1}$ (energy flux of $(8.9\pm2.3) \times 10^{-13}$ erg cm$^{-2}$ s$^{-1}$).
The 0.1--300 GeV average spectrum of the $\gamma$-ray signal is shown in Figure \ref{spectrum}. And, the data analysis results are listed in Table 1. We note that the $\gamma$-ray signal has been recently reported in the Fermi-LAT 16-yr Source List (FL16Y; \citealp{2026arXiv260222148B}) and was named FL16Y J0907.6+4117. In FL16Y, the $\gamma$-ray signal has a position of [R.A.= 136.923$\degr$, Decl.= 41.291$\degr$], a statistical significance of $\sim4.2\sigma$, a photon spectral index of $\Gamma_{\gamma} = 2.2\pm0.2$, and energy flux of $(9.1\pm2.5) \times 10^{-13}$ erg cm$^{-2}$ s$^{-1}$. Our results are in good agreement with those reported in FL16Y (\citealp{2026arXiv260222148B}). However, no low-energy counterpart is reported in FL16Y.

To further investigate this, we perform an additional analysis using the 3--300 GeV observation data, following the procedure adopted in \cite{2022ApJS..260...53A}. We use all events with zenith angles less than 105. The $\gamma$-ray signal still can be found near GB6 J0906+4124. Then, we also estimate the best-fit position of the $\gamma$-ray signal. The best-fit position of [R.A.= 136.951$\degr$, Decl.= 41.311$\degr$] is obtained. The position angle of uncertainty ellipse is 179.373$\degr$, the corresponding $1\sigma$ uncertainty along major axis and minor axis of uncertainty ellipse are 0.048$\degr$ and 0.040$\degr$, respectively. We also show the best-fit position and its $1\sigma$ and $3\sigma$ error ellipses in Figure \ref{TSmap}(b). The radio position of GB6 J0906+4124 is offset from the best-fit position by 0.199$\degr$ and lies outside the $3\sigma$ error ellipse, as seen in Figure \ref{TSmap}(b). We obtain TS = 22.0 ($\sim4.2\sigma$)\footnote{TS = 22.0 with 2 degrees of freedom is obtained before the best-fit localization, corresponds to an estimated statistical significance of $\sim4.2\sigma$ (see \citealp{1996ApJ...461..396M}).}, $\Gamma_{\gamma} = 2.2\pm0.2$, and $F_{\rm 3-300 GeV} = (2.6\pm0.8) \times 10^{-11}$ ph cm$^{-2}$ s$^{-1}$ (energy flux of $(4.5\pm1.8) \times 10^{-13}$ erg cm$^{-2}$ s$^{-1}$) for the $\gamma$-ray signal. The data analysis results are also listed in Table 1.

Therefore, based on the positional information alone, the association between the $\gamma$-ray signal and GB6 J0906+4124 \textbf{cannot} be definitively established. In this scenario, we add the $\gamma$-ray signal as a background source, and then estimate the 0.1--300 GeV photon flux upper limit at 95\% confidence level of GB6 J0906+4124 by fixing $\Gamma_{\gamma}$ to 2.0. We obtain $\rm TS \sim 0$, photon flux upper limit of $1.70 \times 10^{-10}$ ph cm$^{-2}$ s$^{-1}$ for GB6 J0906+4124, as listed in Table 2.

\subsection{Data Analysis Results for Other CSOs}

Table 2 also lists the data analysis results of the remaining 72 CSOs. Among the remaining 72 CSOs, the TS values of 69 CSOs are less than 10. For these CSOs, we only derive the 0.1--300 GeV photon flux upper limits at 95\% confidence level by fixing $\Gamma_{\gamma}$ to 2.0. The photon flux upper limits and corresponding luminosity upper limits of these 69 CSOs are given in Table 2. 
The $\gamma$-ray excesses (10 $\leqslant$ TS < 25) detected toward the three CSOs (B3 0402+379, JVAS J1311+1658, and PKS 1732+094) are regarded as marginal detections. 
For CSO B3 0402+379, the $\gamma$-ray signal has a TS value of only 13. The derived photon spectral index and photon flux are $2.56\pm0.20$ and ($4.56\pm1.72$) $\times 10^{-9}$ ph cm$^{-2}$ s$^{-1}$, respectively.
The $\gamma$-ray signal from CSO JVAS J1311+1658 has low significance (TS = 14), we derive its photon spectral index and photon flux and obtain $2.86\pm0.25$ and ($3.31\pm1.15$) $\times 10^{-9}$ ph cm$^{-2}$ s$^{-1}$, respectively.
CSO PKS 1732+094 exhibits a weak $\gamma$-ray signal (TS = 18), yielding a photon spectral index and photon flux of $2.57\pm0.17$ and ($4.30\pm1.72$) $\times 10^{-9}$ ph cm$^{-2}$ s$^{-1}$ respectively. The luminosities of these three CSOs are also listed in Table 2.

\section{Discussion on the $\gamma$-ray Signal Near GB6 J0906+4124}

As presented in Section 3, the radio position of GB6 J0906+4124 is outside the $3\sigma$ error ellipses of the best-fit positions of the $\gamma$-ray signal in both 0.1--300 GeV and 3--300 GeV bands. If we rely solely on positional coincidence, GB6 J0906+4124 would not be considered a robust counterpart.

We examine sources located within the $3\sigma$ error ellipses of the best-fit positions for both 0.1--300 GeV and 3--300 GeV bands. Apart from a radio source named NVSS J090706+411426, a Seyfert galaxy, and a few quasars, most of the sources are galaxies and stars. With the exception of NVSS J090706+411426, neither the Seyfert galaxy nor the quasars have radio observations. 
And, NVSS J090706+411426 has only radio data available and lacks multi-wavelength data. Nevertheless, it remains a possible counterpart. However, given the lack of multi-wavelength information to establish a robust association, we therefore broaden our search to the 68\% containment radius of Fermi-LAT average PSF at 3 GeV ($\sim0.3\degr$).
In Figure \ref{TSmap}(b), the 68\% containment radius is also shown. A total of 160 sources lie within the 68\% containment radius. These include CSO GB6 J0906+4124, 44 quasars, 5 radio galaxies, 3 radio sources, 2 Seyfert galaxies, 1 LINER, and 1 AGN candidate, while the majority of the remaining objects are classified as galaxies or stars. 43 quasars, 2 Seyfert galaxies and 1 LINER lack both radio and X-ray observation data. Given the lack of radio and X-ray data, these sources cannot be meaningfully evaluated as counterparts and are thus not considered further. We list the remaining possible candidates in Table 3, including their classifications, positions, offsets, observed bands, and 1.4 GHz radio flux densities.

We additionally employ the \textit{gtsrcid} tool to compute the association probabilities for these candidate sources located within the 68\% containment radius by considering the Faint Images of the Radio Sky at Twenty-centimeters (FIRST) survey catalog (\citealp{2012yCat.8090....0B}). The \textit{gtsrcid} tool is an application that finds counterparts for a list of detected sources using a catalog of potential counterparts (\citealp{2010ApJS..188..405A}). With its default configuration, \textit{gtsrcid} tool computes the association probability based solely on positional coincidence. The resulting probability between GB6 J0906+4124 and the $\gamma$-ray signal is 55.7\%, which is listed in Table 3. The resulting probabilities between other candidate sources and the $\gamma$-ray signal are also presented in Table 3.

As seen in Table 3, 11 candidate sources are listed. Four sources have only radio observations, four sources have radio-to-optical data, one source has radio-to-IR data, and only two sources (NVSS J090652+411519 and GB6 J0906+4124) are covered from radio to UV. 
GB6 J0906+4124 has a positional offset of 0.199$\degr$ and an association probability of 55.7\%, ranking 7th. Nonetheless, GB6 J0906+4124 is still a plausible counterpart.
First, GB6 J0906+4124 is a core-dominated source (\citealp{2018ApJ...863..155C, 2021MNRAS.506.1609C}), and it is the only source that has been previously classified as a blazar (\citealp{2013MNRAS.430.2464M, 2023Ap.....66...11A}), a class that dominates the Fermi-LAT extragalactic $\gamma$-ray sky. 
Second, it is one of only two candidate sources with radio-to-UV spectral coverage. 
In contrast, six sources with smaller offsets and higher association probabilities are all either unclassified radio sources or radio galaxies with limited observation data or lower 1.4 GHz flux densities.

We also note the presence of the blazar 4C +41.18 at an angular distance of 0.33$\degr$ from the 3--300 GeV best-fit position of the $\gamma$-ray signal. This source is a known blazar and thus is physically a plausible type of $\gamma$-ray emitter. Its available multi-wavelength coverage is limited to radio-to-optical bands, but it has a 1.4 GHz flux density one order of magnitude higher than that of GB6 J0906+4124. However, its larger offset compared to GB6 J0906+4124 makes it a less likely counterpart based on positional grounds. Nevertheless, we cannot completely exclude the possibility that the $\gamma$-ray signal originates from 4C +41.18.

\section{Summary}
\label{sec:Con}

We utilized 16 yr Fermi-LAT observation data to search for $\gamma$-ray emitting CSOs. 
The search sample consists of 73 sources cataloged as bona fide CSOs  (\citealp{2024ApJ...961..240K}).
Among these sources, a significant $\gamma$-ray signal near CSO GB6\,J0906+4124 is detected with TS = 28.7 ($\sim4.9\sigma$) in the 0.1--300 GeV band. Data analysis in both the 0.1--300 GeV and 3--300 GeV bands revealed that the radio position of GB6\,J0906+4124 lies outside the $3\sigma$ error ellipse of the best-fit position for this $\gamma$-ray signal. A search for possible counterparts within the 68\% containment radius of Fermi-LAT average PSF at 3 GeV reveals 11 sources, among which GB6 J0906+4124 remains a plausible candidate due to its classification and broad spectral coverage. Furthermore, based on the 0.1--300 GeV observation data, we found $\gamma$-ray marginal detections ($10 \leqslant \mathrm{TS} < 25$) for three additional CSOs, i.e., B3 0402+379, JVAS J1311+1658, and PKS 1732+094. For the remaining non-detected sources, we estimated their 0.1--300 GeV photon flux upper limits, providing the systematic constraints on the $\gamma$-ray emission from CSOs.

Our results suggest that GeV $\gamma$-ray emission from CSOs is not common, or that such emission is typically below the current Fermi-LAT detection sensitivity. Deeper multi-wavelength observations are needed to confirm the origin of the $\gamma$-ray signal near GB6 J0906+4124 and to further understand the high-energy properties of CSOs.

\acknowledgments
 
We thank the anonymous referee for valuable suggestions. This work utilized the NASA/IPAC Extragalactic Database funded by the National Aeronautics and Space Administration and operated by the California Institute of Technology. This work acknowledges the general academic support provided by Professor Jin Zhang during the initial phase of the study at School of Physics, Beijing Institute of Technology.

\clearpage

\bibliography{reference}
\bibliographystyle{aasjournal}

\clearpage

\begin{deluxetable}{lccccc}
\tabletypesize{\scriptsize} 
\tablecolumns{7} 
\tablewidth{0pc}
\tablecaption{Fermi-LAT analysis results of the $\gamma$-ray signal near GB6 J0906+4124.}
\tablehead{\colhead{Energy Range} & \colhead{$R.A.$$^a$} & \colhead{$Decl.$$^a$} & \colhead{TS} & \colhead{$\Gamma_{\gamma}$} & \colhead{Photon Flux} \\
\colhead{[GeV]} & \colhead{[deg]} & \colhead{[deg]} & \colhead{} & \colhead{} & \colhead{[ph cm$^{-2}$ s$^{-1}$]}}
\startdata
0.1--300&136.905&41.315&28.7&$2.1\pm0.2$&$(9.7\pm5.6) \times 10^{-10}$\\
3--300&136.951&41.311&22.0&$2.2\pm0.2$&$(2.6\pm0.8) \times 10^{-11}$\\
\enddata 
\tablenotetext{a}{The best-fit positions for 0.1--300 GeV and 3--300 GeV bands.}
\end{deluxetable}

\begin{startlongtable}
\begin{deluxetable}{lcccccc}
\tabletypesize{\scriptsize} 
\tablecolumns{7} 
\tablewidth{0pc}
\tablecaption{Fermi-LAT analysis results of CSOs.}
\tablehead{\colhead{J2000 Name} & \colhead{Common Name} & \colhead{$z$} & \colhead{$\Gamma_{\gamma}$} & \colhead{$F_{\rm 0.1-300~GeV}$} & \colhead{log$L_{\rm 0.1-300~GeV}$} & \colhead{TS}\\
\colhead{} & \colhead{} & \colhead{} & \colhead{} & \colhead{[$\times 10^{-10}$ ph cm$^{-2}$ s$^{-1}$]} & \colhead{[erg s$^{-1}$]} & \colhead{}}
\startdata
J0000+4054&B3 2358+406&&2.0&<1.37&&<10\\
J0003+4807&JVAS J0003+4807&&2.0&<3.68&&<10\\
J0029+3456&B2 0026+34&0.517&2.0&<7.31&<44.99&<10\\
J0111+3906&0108+388&0.66847&2.0&<1.49&<44.57&<10\\
J0119+3210&B2 0116+31&0.0602&2.0&<5.55&<42.78&<10\\
J0132+5620&JVAS J0132+5620&&2.0&<6.24&&<10\\
J0150+4017&B3 0147+400&&2.0&<1.95&&<10\\
J0204+0903&JVAS J0204+0903&&2.0&<1.68&&<10\\
J0237+4342&B3 0233+434&&2.0&<0.87&&<10\\
J0402+8241&JVAS J0402+8241&0.065568&2.0&<3.56&<42.67&<10\\
J0405+3803&B3 0402+379&0.05505&2.56$\pm$0.20&45.56$\pm$17.17&43.15$\pm$0.15&13\\
J0425--1612&PKS 0423-163&&2.0&<1.10&&<10\\
J0427+4133&B3 0424+414&&2.0&<2.21&&<10\\
J0440+6157&GB6 J0440+6158&&2.0&<3.25&&<10\\
J0706+4647&B3 0703+468&&2.0&<5.65&&<10\\
J0713+4349&B3 0710+439&0.518&2.0&<1.25&<44.22&<10\\
J0735--1735&PKS 0733-17&&2.0&<8.94&&<10\\
J0741+2706&B2 0738+27&0.772139&2.0&<4.46&<45.20&<10\\
J0754+5324&JVAS J0754+5324&0.84&2.0&<3.29&<45.16&<10\\
J0832+1832&PKS 0829+18&0.154&2.0&<2.95&<43.38&<10\\
J0855+5751&JVAS J0855+5751&0.025998&2.0&<3.78&<41.86&<10\\
J0906+4124&GB6 J0906+4124&0.0273577&2.0&<1.70&<41.87&<10\\
J0909+1928&MRK 1226&0.027843&2.0&<4.52&<42.00&<10\\
J0943+1702&JVAS J0943+1702&1.601115&2.0&<1.71&<45.57&<10\\
J1011+4204&B3 1008+423&&2.0&<3.65&&<10\\
J1025+1022&NVSS J102544+102231&0.045805&2.0&<3.43&<42.32&<10\\
J1035+5628&JVAS J1035+5628&0.46&2.0&<2.00&<44.30&<10\\
J1042+2949&B2 1039+30B&0.61&2.0&<1.76&<44.55&<10\\
J1111+1955&PKS 1108+201&0.299&2.0&<6.81&<44.39&<10\\
J1120+1420&PKS 1117+146&0.362&2.0&<1.42&<43.91&<10\\
J1135+4258&B3 1133+432&1.00&2.0&<1.78&<45.08&<10\\
J1158+2450&PKS 1155+251&0.203&2.0&<6.13&<43.96&<10\\
J1159+5820&VERA J1159+5820&1.27997&2.0&<1.57&<45.29&<10\\
J1204+5202&GB6 J1204+5202&0.01&2.0&<2.89&<40.90&<10\\
J1205+2031&NGC 4093&0.02378857&2.0&<3.03&<41.69&<10\\
J1227+3635&B21225+36&1.975&2.0&<4.04&<46.17&<10\\
J1234+4753&JVAS J1234+4753&0.373082&2.0&<5.07&<44.49&<10\\
J1244+4048&B3 1242+410&0.813586&2.0&<6.26&<45.40&<10\\
J1247+6723&JVAS J1247+6723&0.107219&2.0&<1.50&<42.74&<10\\
J1254+1856&CRATES J1254+1856&0.1145&2.0&<4.61&<43.29&<10\\
J1311+1658&JVAS J1311+1658&0.081408&2.86$\pm$0.25&33.08$\pm$11.50&43.26$\pm$0.12&14\\
J1313+5458&JVAS J1313+5458&0.613&2.0&<2.35&<44.67&<10\\
J1326+3154&DA 344&0.36801&2.0&<2.24&<44.12&<10\\
J1335+5844&JVAS J1335+5844&0.59&2.0&<2.21&<44.61&<10\\
J1347+1217&PKS B1345+125&0.121&2.0&<1.89&<42.96&<10\\
J1400+6210&1358+625&0.431&2.0&<0.97&<43.92&<10\\
J1407+2827&OQ +208&0.077&2.0&<3.72&<42.83&<10\\
J1413+1509&JVAS J1413+1509&0.35&2.0&<1.80&<43.97&<10\\
J1414+4554&B3 1412+461&0.186&2.0&<3.32&<43.61&<10\\
J1434+4236&B3 1432+428B&0.452&2.0&<1.38&<44.12&<10\\
J1440+6108&VIPS J14402+6108&0.445365&2.0&<1.19&<44.04&<10\\
J1443+4044&B3 1441+409&0.15&2.0&<1.15&<42.94&<10\\
J1508+3423&VV 059a&0.045565&2.0&<1.09&<41.82&<10\\
J1511+0518&JVAS J1511+0518&0.084&2.0&<2.76&<42.78&<10\\
J1559+5924&JVAS J1559+5924&0.0602&2.0&<5.43&<42.77&<10\\
J1602+5243&4C +52.37&0.105689&2.0&<4.14&<43.17&<10\\
J1609+2641&CTD 93&0.473&2.0&<1.94&<44.32&<10\\
J1645+2536&PKS 1642+25&0.588&2.0&<1.86&<44.53&<10\\
J1734+0926&PKS 1732+094&0.735&2.57$\pm$0.17&43.00$\pm$17.22&45.66$\pm$0.13&18\\
J1735+5049&CGRaBS J1735+5049&0.835&2.0&<1.63&<44.85&<10\\
J1816+3457&B2 1814+34&0.245&2.0&<5.37&<44.09&<10\\
J1826+1831&JVAS J1826+1831&&2.0&<4.94&&<10\\
J1826+2708&B2 1824+27&&2.0&<1.37&&<10\\
J1915+6548&JVAS J1915+6548&0.486&2.0&<3.53&<44.61&<10\\
J1928+6815&JVAS J1928+6814&&2.0&<1.74&&<10\\
J1939--6342&PKS 1934-63&0.183&2.0&<1.07&<43.10&<10\\
J1944+5448&S4 1943+54&0.263&2.0&<4.30&<44.06&<10\\
J1945+7055&S5 1946+70&0.101&2.0&<6.64&<43.33&<10\\
J2022+6136&S4 2021+61&0.2266&2.0&<0.91&<43.24&<10\\
J2203+1007&JVAS J2203+1007&1.005&2.0&<1.01&<44.84&<10\\
J2327+0846&NGC 7674&0.02892&2.0&<5.00&<42.08&<10\\
J2347--1856&PKS 2344-192&&2.0&<1.70&&<10\\
J2355+4950&TXS 2352+495&0.23831&2.0&<1.03&<43.35&<10\\
\enddata 
\end{deluxetable}
\end{startlongtable}

\begin{deluxetable*}{lccccccc}
\tabletypesize{\scriptsize} 
\tablecolumns{8} 
\tablewidth{0pc}
\tablecaption{Candidate Sources.}
\tablehead{\colhead{Name} & \colhead{Type$^a$} & \colhead{$R.A.$$^b$} & \colhead{$Decl.$$^b$} & \colhead{Offset$^c$} & \colhead{Observed Bands} & \colhead{Flux Density$^d$} & \colhead{Probability} \\
\colhead{} & \colhead{} & \colhead{[deg]} & \colhead{[deg]} & \colhead{[deg]} & \colhead{} & \colhead{[Jy]}}
\startdata
NVSS J090706+411426&Rad&136.777&+41.241&0.148&radio&1.05E-1&61.2\%\\
NVSS J090752+410927&RG&136.971&+41.155&0.157&radio-optical&7.90E-3&60.3\%\\
NVSS J090751+412812&Rad&136.965&+41.468&0.157&radio&2.03E-1&60.3\%\\
NVSS J090821+412559&RG&137.090&+41.436&0.163&radio&2.30E-3&59.8\%\\ 
FIRST J090700.3+411215&AG?&136.751&+41.204&0.184&radio-IR&5.25E-3&57.5\%\\
NVSS J090652+411519&QSO&136.716&+41.252&0.186&radio-UV&3.50E-3&57.2\%\\
GB6 J0906+4124&CSO/Bla&136.720&+41.408&0.199&radio-UV&5.97E-2&55.7\%\\
FIRST J090710.2+412953&RG&136.793&+41.498&0.221&radio-optical&4.36E-3&52.7\%\\
NVSS J090844+410606&RG&137.187&+41.103&0.274&radio&6.67E-2&44.5\%\\
4C 41.19&Rad&136.887&+41.583&0.276&radio-optical&1.40&44.1\%\\
NVSS J090619+411436&RG&136.583&+41.245&0.284&radio-optical&1.67E-2&42.6\%\\
\enddata 
\tablenotetext{a}{Source classification obtained from the SIMBAD Astronomical Database: Rad---radio source; RG---radio galaxy; AG?---AGN Candidate; QSO---quasar; Bla---Blazar.}
\tablenotetext{b}{Source positions are obtained from the FIRST survey catalog (\citealp{2012yCat.8090....0B}).}
\tablenotetext{c}{The angular distances between the source position and the 3--300 GeV best-fit position of the $\gamma$-ray signal.}
\tablenotetext{d}{1.4 GHz flux densities obtained from NED.}
\end{deluxetable*}

\end{CJK}
\end{document}